# Time Series Analysis of Computer Network Traffic in a Dedicated Link Aggregation

G. Millán, G. Lefranc, R. Osorio-Comparán, and V. Lomas-Barrie

**Abstract**— Fractal behavior and long-range dependence are widely observed in measurements and characterization of traffic flow in high-speed computer networks of different technologies and coverage levels. This paper presents the results obtained when applying fractal analysis techniques on a time series obtained from traffic captures coming from an application server connected to the Internet through a high-speed link. The results obtained show that traffic flow in the dedicated high-speed network link have fractal behavior when the Hurst exponent is in the range of 0.5, 1, the fractal dimension between 1, 1.5, and the correlation coefficient between –0.5, 0. Based on these results, it is ideal to characterize both the singularities of the traffic and its impulsiveness during a fractal analysis of temporal scales. Finally, based on the results of the time series analyses, the fact that the traffic flows of current computer networks exhibit fractal behavior with a long-range dependency is reaffirmed.

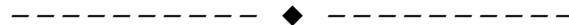

## 1 INTRODUCTION

TRAFFIC flow are useful for having a under-standing of traffic on a computer network, providing a measurement of traffic and to know what hosts are talking on the network, with details of addresses, volumes and types of traffics and protocols. This knowledge can be useful for troubleshooting, detecting security incidents, and planning and network design [1].

Performance traffic models require to be accurate and to have the ability to capture the statistical characteristics of the actual traffic on the network. Many traffic models have been developed based on traffic measurement data.

It is necessary to analyse network traffic the communications on computer network to find evidence of security threat invasion; to decide the Quality of Service level; and others issues such as data transmitted through outdated switches, routers, servers, and Internet exchanges that can cause bottlenecks. The result is network congestion. If it is detected suspicious traffic, the team is alerted to the issue in real-time.

High-speed computer networks provide high-speed links iff economy of scale; bursty, short holding time traffic; shared-switch distributed-media, no shared-media access; speed-distance-transmission size tradeoff.

In computer networking, the term link aggrega-tion refers to various methods of combining multi-ple network connections in parallel in order to increase throughput beyond what a single connection could sustain, and to provide redundancy in case one of the links should fail.

Time-series analysis is employed in a network performance monitoring architecture, to provide services for event triggering, alarming, and statistical auditing. One such application is anomaly detection, which can be utilized for performance and security management. Forecasting is also a relevant exercise, where the history of the network behavior and usage is exploited to predict future performance [2].

Fractal behavior and Long-range dependence (LRD) are observed in many phenomena, such as in nature [3]–[8], in financial time series [9], in communication system traffic [10]–[14], and in heart rate time series [15], [16]. This article characterizes the time-series dynamics of traffic flows captured from a high-speed dedicated link connecting an application server and the Internet, by applying fractal analysis considering the following test: Detrended Fluctuation Analysis (DFA), Power Spectral Analysis (PSA), and Time-Scale Analysis (TSA).

There are two modeling streams: a conventional one, which bases its assumptions on generally Markov processes, and another self-similar one, which accepts the LRD as an inherent singularity of data traffic flows.

The research related to traffic self-similarity can be classified into four categories: measurement-based traffic modelling, physical modelling, queuing analysis and traffic control as well as resource provisioning [17].

In [18] reports the results from the analysis of the computer network traffic using the statistical self-similarity factor. The analyzed traffic has a self-similar nature to the degree of self-similarity in the range of 0.5 to 1.

Fractal behavior and long-term dependence are widely observed in the measurements and characterization of traffic flow in high-speed computer networks of different technologies and coverage levels [1]. It is proposes to obtain the fractal behavior of network traffic data based on topology, to reduce the complexity in the network [19].

Several approaches have explored to calculate the fractal dimension of a subset with respect to a fractal structure. A discrete models of fractal dimension to explore the complexity of discrete dynamical systems [20].

A simple and fast technique of multifractal traffic modeling has been proposed and a method of fitting model to a given traffic trace. A comparison of simulation results obtained for an exemplary trace, multifractal

- *G. Millán is with the DIE, USACH, E-mail: ginno.millan@usach.cl.*
- *G. Lefranc is is with the EIE, PUCV.*
- *R. Osorio-Comparán is with the DISCA, IIMAS, UNAM.*
- *V. Lomas-Barrie is with the DISCA, IIMAS, UNAM.*

model and Markov Modulated Poisson Process (MMPP) models has been performed [15], [21].

In the paper presents the results obtained when applying fractal analysis techniques on a time series found from traffic captures coming from an application server connected to the Internet through a high-speed link.

The data analyzed correspond to the size of traffic frames of the central online applications server at Universidad de Santiago de Chile, which serves 20000 users connected online through the internet. This article analyzes two different types of traffic flows, **SERV-1** and **SERV-2**. **SERV-1** is the temporary series of frame sizes that are transferred to the server from the Internet and **SERV-2** is the temporary series of frame sizes that are transferred from the server to the Internet. These traffic flows play an important role in determining the degree of smooth access to the corresponding application server and therefore the Quality of Service provided to users and the Quality of Experience that users perceive [22], [23].

The traffic bursts over extensive periods reveal that the traffic flows under study are identified with a completely different nature from those predicted by a classic Poisson model related to the traffic flows of the old telephone system. For this reason, this research focuses on applying a broad battery of fractal analysis that reaffirms that traffic flows in current high-speed computer networks are fractal with LRD, regardless of their sources such as device requesting services [24]. This research is about a high-speed dedicated link and an on-line application server. It should be noted that the time series come from the capture of packets on said link and therefore can be generalized in terms of the presence of traffic from both the Internet and from within the corporate network of the Universidad de Santiago de Chile.

This paper presents the results obtained when applying fractal analysis techniques on a time series obtained from traffic captures coming from an application server connected to the Internet through a high-speed link. The results obtained show that traffic flow in the dedicated high-speed network link have fractal behavior. Based on these results, it is ideal to characterize both the singularities of the traffic and its impulsiveness during a fractal analysis of temporal scales. Based on the results of the time series analyses, the traffic flows exhibit fractal behavior with a long-range dependency.

The article is structured as follows. First, we present the general aspects of Fractal Processes (FP), followed by the key aspects of DFA, PSA, and TSA. Then, the main results obtained are presented and their validity is discussed. Finally, the main aspects of the research and the conclusions are presented.

## 2 THEORETICAL FOUNDATION

### 2.1 Fractal Processes

A Fractal Processes (FP) is characterized by having a non-integrer dimension, $D$. Also, a FP has two characteristics inherent to its phenomemology 1) a FP is like itself even at different observation scales. This property is known as invariance at the scale. The Self-similarity exists when the process exhibits a similar behavior under isotropic scaling and 2) a FP consists of a complex internal structure and shows the same behavior even at different magnification scales, i.e. FP has a self-similar hierarchical structure [25].

Due to the scale invariance, there is a power-law behavior between two parameters in a FP that is governed by the relationship $f(x) \propto x^c$, where $f(x)$ is a function of a study object and $c$ is a constant.

In [20] they estimate $D$ based on the power-law behavior expressed by the above expression. Moreover from the definition of fractional Brownian motion (fBm), these fBm processes must be governed by [26]

$$B_H(t) = [\tau(H+0.5)]^{-1}\left(\int_{-\infty}^{0}[(t-s)^{H-0.5}-(-s)]dB(s) + \int_{0}^{t}(t-s)^{H-0.5}dB(s)\right), \quad (1)$$

where $0 < H < 1$ is the Hurst exponent of the fBm process.

Additionally, $B_H(t)$ satisfies

$$E[B_H(t)] = 0, \quad (2)$$

$$E[B_H^2(t)] \sim t^{2H}, \quad (3)$$

$$E[B_H(t)B_H(s)] = 0.5(|t|^{2H} + |s|^{2H} - |t-s|^{2H}). \quad (4)$$

From (4) the correlation coefficient, $\rho$, between the $B_H(t)$ successive increments can be written in the form

$$\rho = \left\langle \frac{-B_H(-t)B_H(t)}{B_H^2(t)} \right\rangle, \quad (5)$$

where

- If $t = t_0$, then $B_H(t = t_0)$,
- If $t = -t$, then $B_H(t = -t) = B_H(-t)$, and
- $B_H(t) = -B_H(t)$, for all $t$.

Therefore, we have

$$\rho = 2^{2H-1} - 1. \quad (6)$$

Then, be $y(t)$ a FP with a Hurst exponent given by $H$ and then for an arbitrary process with

$$y(ct) \triangleq c^H y(t), \quad c > 0, \quad (7)$$

is also a FP with the same statistical distributions than the $y(t)$ process, and in which it is verified that $D$ is given by the expression [20]

$$D = 2 - H. \quad (8)$$

Table 1 shows the relationships between $H$, $D$, $\rho$, and FP behavior.

TABLE 1
INTERVALS OF *H* AND *D* VALUES AND THEIR ASSOCIATED PROCESSES

| H | D (8) | ρ (6) | FP Behavior |
|---|---|---|---|
| > 0.5 | < 1.5 | Positive | Persistent |
| = 0.5 | = 1.5 | Random | fBm |
| < 0.5 | > 1.5 | Negative | Non-persistent |

## 2.2 Power-Spectral Analysis (PSA)

Time series can be described in the time-domain by $x(t)$, but can also be described in the frequency domain by Fourier Transform (FT), $X(\omega)$, where $\omega$ angular frequency.

The autocorrelation function of a non-stationary time series $x(t)$, is given by

$$R_{xx}(t+\tau) = \int_{-\infty}^{\infty} E[x(t)x(t+\tau)]dt, \qquad (9)$$

The FT of (9) is the same as $|X(\omega)|^2$ therefore the power-spectral density (PSD), $S(\omega)$, can be written as

$$S(\omega) \triangleq |X(\omega)|^2. \qquad (10)$$

Using the Wiener-Khintchine theorem, the time series PSD can be expressed as the FT of (9) as follows

$$S_{xx}(\omega) = \int_{-\infty}^{\infty} R_{xx}(\tau) e^{-j\omega\tau} d\tau. \qquad (11)$$

The power-spectral function provides an important parameter to characterize persistence in time series. For a fractal time series, its power-spectral function [20] obeys the frequency-based power-law behavior and is given by the expression

$$S_m(\omega) \cong \omega_m^{-\beta}, \text{ with } m = 1,2,...,N/2, \qquad (12)$$

where $\omega_m = n/N$; $N$ the length of the time series and $\beta$ the spectral-exponent that characterizes series persistency.

The relationship between $\beta$, $H$, and $D$ is given by [20]

$$\beta = 2H + 1 = 5 - 2D. \qquad (13)$$

This expression allows to obtain the value of $\beta$ using the least-squares method on the adjustment curves of $H$ or $D$.

The PSA method only provides the global value of $H$ from the FT using a harmonic function. However, it is traditional in fractal analysis for its simplicity to obtain based on an estimate of the real $H$ value [27].

## 2.3 Detrended Fluctuation Analysis (DFA)

The DFA was widely used to determine the scaling properties of self-similar processes and to determine LRD on noisy and non-stationary time series. In general, this type of analysis is used to estimate the fluctuation of the RMS (Root-Mean-Square) of series with and without a trend (this latter case is a variant the RMS analysis of the processes based on the theory of random walks [28]), and also because it can detect LRD.

The mathematical form of a time series $Y(i)$, is given according to [29] by

$$Y(i) = \sum_{k=1}^{i}(x_k - \langle x \rangle), \text{ with } i = 1,2,...,N, \qquad (14)$$

where $x_k$ is the $k$th sequence of the time series of length $N$, and $\langle x \rangle$ is its average.

Then the series $Y(i)$ given by (14) is regrouped in $N_s \equiv$ Int $(Ns^{-1})$ on non-overlapping segments of equal length, $s$, as shown in Fig. 1, a process which is also known as aggregation.

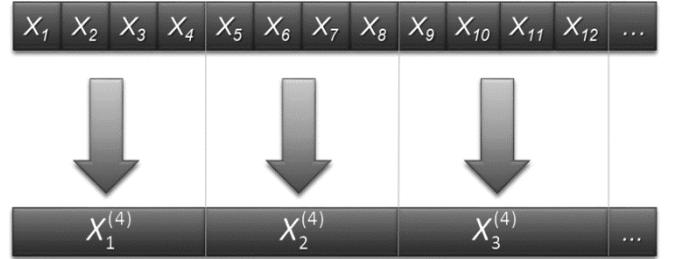

Fig. 1. Aggregation process of non-overlapping segments for a time series.

As it often happens, the lengths of the time series are not a multiple of the time-scale, $s$, so a short part of it remains at the end of the aggregate series.

To solve this problem, the same procedure is repeated but this time starting from the opposite end and analyzing the part that will remain at the beginning of the aggregate series; therefore, the total number of segments is $2N_s$.

After the aggregate time series composed of $N_s$ segments of length $s$ have been obtained, an optimal adjustment line is projected using the least-squares method in each series to obtain the local tendency of each segment that composes it.

The deviation of each time series is obtained from the subtraction of the line of best fit of the minimum squares and the variance which is calculated by the expression

$$F^2(s,v) \equiv \frac{1}{s}\sum_{i=1}^{s}\{Y[(v-1)s+i] - y_v(i)\}^2, \qquad (15)$$

for each segment $v$, with $v = 1,..., N_s$, and

$$F^2(s,v) \equiv \frac{1}{s}\sum_{i=1}^{s}\{Y[N-(v-N_s)s+i] - y_v(i)\}^2, \qquad (16)$$

for each segment $v = N_s+1,..., 2N_s$, where $y_v(i)$ corresponds to the best adjustment line obtained by using the least-squares method in segment $v$.

The last step of the DFA analysis is to obtain the average of all segments of each time series disaggregated to find the function given by

$$F(s) \equiv \frac{1}{2N_s} \sum_{v=1}^{2N_2} F^2(s,v), \quad (17)$$

where $F(s)$ increases as $s$ increases and is defined only for segments of length $s \geq 4$. Therefore, the previous steps are repeated several times to obtain a data set of $F(s)$ versus $s$, where the slope of the curve obtained from that graph represents the scaling exponent $\alpha$ if the series is correlated according to a long-range power-law.

Therefore, $F(s)$ and $s$ are related by the power-law

$$F(s) \sim s^\alpha. \quad (18)$$

Table 2 relates the scaling exponent $\alpha$ to different types of processes.

TABLE 2
RELATIONSHIP BETWEEN $\alpha$ AND PROCESS TYPES

| $\alpha$ Interval | Process Type |
|---|---|
| 0 < $\alpha$ < 0.5 | Power-law anti-correlation |
| $\alpha$ = 0.5 | White noise |
| 0.5 < $\alpha$ < 1 | Long-range power-law correlation |
| $\alpha$ = 1 | 1/$f$ process |
| $\alpha$ > 1 | fBm process |

## 2.4 Time-Scale Analysis (TSA)

The methods presented in the previous sections are based on the development of a linear log-log type graph that only outputs a unique $H$ value. These methods are insufficient when estimating the locally time-dependent Hurst exponent, $H(t)$ [30], [31].

The Wavelet Transform approach results in a powerful mathematical tool that serves for both the hierarchy of a FP and spatial distribution of the singularities of the fractal measurements. In this research only the Continuous Wavelet Transform (CWT) is considered for temporal scales analysis to estimate $H(t)$ [32].

It should be noted that in the literature $H$ is a global (also called general) Hurst exponent, and $H(t)$ as a local Hurst exponent [33], [34].

So, the CWT is defined as [35]

$$W_x(t,a,\varphi) = \int_{-\infty}^{\infty} x(s)\varphi_{t,a}^*(s)ds, \quad (19)$$

where $\varphi^*$ is the conjugate complex of $\varphi$ function, that for different observations scales is defined as

$$\varphi_{t,a}(s) = |a|^{0.5} \varphi[(s-t)/a], \quad (20)$$

where $a$ is the scale-parameter and $a \propto \omega - 1$.

In this research the Morlet Wavelet is used for the TSA and its scalogram is defined as

$$E_x = \int_{-\infty}^{\infty}\int_{-\infty}^{\infty} |W_x(t,a,\varphi)|^2 a^2 dt da, \quad (21)$$

where $E_x$ is the energy of function $x$ [36].

Therefore, a scalogram is an energy distribution function of a signal in a time-scale plane associated with $a^2dt\,da$. Concerning the above, in general, any time series is a representation of a signal. Thus, considering time series with uniform $H$ can be described as [37]

$$|x(s) - x(t)| \leq |s-t|^H, \text{ with } c \in \mathbb{R}. \quad (22)$$

Applying CWT for $x(t)$ in (22)

$$|W_x(t,a,\varphi)| \leq |a|^{H+0.5} \int_{-\infty}^{\infty} |t|^H |\varphi(t)| dt, \quad (23)$$

and the scalogram for this time series is given by [37]

$$\Omega(t,a) \equiv |W_x(t,a)|^2 \approx |a|^{2H(t)+1}, \text{ when } a \to 0. \quad (24)$$

Based on (24), it is possible to estimate $H(t)$ and write $H$ as follows

$$H = \frac{1}{T}\int_0^T H(t)dt. \quad (25)$$

Thus, the TSA provides both $H$ and $H(t)$.

Therefore TSA is a more powerful mathematical tool compared to PSA and DFA in FP analysis since most of traffic flow processes exhibit multifractal scaling behaviors and it is possible to characterize them with the fluctuations of $H$ described by $H(t)$.

# 3 FRACTAL ANALYSIS DEVELOPMENT

## 3.1 Preliminary

The test scenario is presented in the following figure

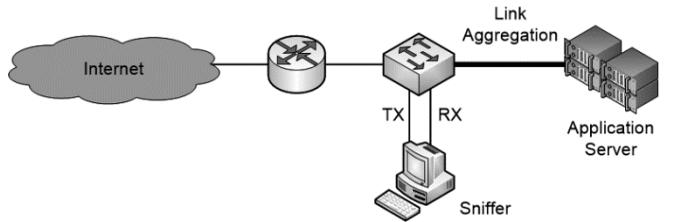

Fig. 2. Network traffic testing scenario.

## 3.2 Fractal Analysis

The spectral exponent ($\beta$), $H$, $D$, and $\rho$ of the **SERV-1** and **SERV-2** time series estimated with the PSA method are tabulated in Table 3. It is emphasized that the spectral exponent is defined in (12) and is related to $H$ and $D$ by means (13); it stands out that $\rho$ is related to $H$ through (6). The results clearly show that the **SERV-1** and **SERV-2** time series exhibit fractal behavior with LRD that agrees with the theory.

To test the accuracy of the DFA algorithm which used in this research, the algorithm is used to calculate the scaling exponent of three known scaling exponent generated signals, wich are Brownian motion, persistence power-law, and anti-persistence power-law processes with $H$ = 0.50, $H$ = 0.80, and $H$ = 0.20 [37], respectively.

The results are shown in Table 4.

The results show that the scaling exponents obtained are consistent with the $H$ for the three generated series, which verifies that the DFA method carried out in the fluctuation analysis without tendency is assertive to reproduce results.

The scaling exponent ($\alpha$) of **SERV-1** and **SERV-2** series estimated with the DFA method are shown in detail in Table 5.

The results show complete coherence with the theory estaments and that the behavior of the time series under study, responded to a fractal character with LRD trend. The experiment on the scaling exponent reflects that both series respond to a behavior of the fractal type with LRD.

The scalogram allows $H$ and $H(t)$ to be estimated for **SERV-1** and **SERV-2** time series. The results applying the TSA method are summarized in Table 6.

TABLE 3
NUMERICAL EXPERIMENTS FOR **SERV-1** AND **SERV-2** TIME SERIES CONSIDERING $H$, $D$, $\beta$, AND $\rho$

| Time Series | $H$ | $D$ | $\beta$ | $\rho$ |
|---|---|---|---|---|
| SERV-1 | 0.70±0.01 | 1.80±0.01 | 1.60±0.01 | –025±0.01 |
| SERV-2 | 0.71±0.01 | 1.81±0.01 | 1.61±0.01 | –0.24±0.01 |

TABLE 4
$\alpha$ FOR DIFFERENT PROCESSES

| Time Series Type | $H$ | $\alpha$ According to DFA Method | ±$\alpha$ |
|---|---|---|---|
| Brownian Motion | 0.50 | 1.20 | 0.10 |
| Persistence power-law | 0.80 | 1.51 | 0.09 |
| Anti-persistence power-law | 0.20 | 1.80 | 0.03 |

TABLE 5
$\alpha$ FOR **SERV-1** AND **SERV-2** TIME SERIES

| Time Series | $\alpha$ | ±$\varepsilon$ | $\alpha$ | ±$\varepsilon$ | $\alpha$ | ±$\varepsilon$ |
|---|---|---|---|---|---|---|
| SERV-1 | 0.65 | 0.04 | 1.08 | 0.05 | 2.01 | 0.05 |
| SERV-2 | 0.64 | 0.03 | 1.07 | 0.05 | 2.00 | 0.04 |

TABLE 6
GLOBAL HURST EXPONENT, LOCAL HURST EXPONENT IN TERMS OF MINIMUM AND MAXIMUM VALUES, AND $D$ FOR **SERV-1** AND **SERV-2** TIME SERIES

| Time Series | $H$ | Min {$H(t)$} | Max {$H(t)$} | $D$ |
|---|---|---|---|---|
| SERV-1 | 0.32 | –0.49 | 1.48 | 1.68 |
| SERV-2 | 0.27 | –0.26 | 1.15 | 1.73 |

From the results given in Tables 3, 4, 5 and 6, it is shown that the two time series under analysis (**SERV-1** that contains the frame sizes that are transferred to the server from the Internet and **SERV-2** that contains the frame sizes that are transferred from the server to the internet) exhibit fractal characteristics with LRD. It is inferred that the increase of samples for any of both series as a result of the extension of the observation time will not result in a modification of their nature, given that these two series have a behavior with LRD.

Even when the FT uses harmonic basis functions and processes non-stationary signals, the PSA is a good way to start with the initial measurements of non-stationary time series that are suspected to have a fractal nature: as is the case of the time series presented in this research.

Two of the main results obtained are:
1) $H = 0.70 \pm 0.01$ in **SERV-1** *time series*. Result that clearly reveals fractal character with LRD trend.
2) $H = 0.67 \pm 0.01$ in **SERV-2** *time series*. Result that clearly reveals fractal character with LRD trend.

It is interesting to examine the results of the fluctuation analysis without tendency since they show that both time series present the crossing phenomenon characteristic described in [13].

The origin of this phenomenon can be explained by the fact that there are very short periods between a service request and the server's response. This generates a time series for a highly fluctuating uncorrelated process. As time passes, the signals show fluctuations that tend to soften, reflecting the dynamics of every current telecommunications system, resulting in an exponent $\alpha = 1$ associated with a process $1/f$.

The results of TSA show that the considered time series are constitutive of extremely complicated systems that present a time-dependent Hurst exponent which ranges from negative to positive values $-0.50 \leq H(t) \leq 1.50$ for the **SERV-1** series and $-0.30 \leq H(t) \leq 1.15$ for the **SERV-2** series. It is further noted that $H(t)$ for the **SERV-1** series has greater complexity than $H(t)$ for the **SERV-2** series. This difference can explain the following; for **SERV-1**, the data comes from thousands of points distributed on the internet to a server entry port, which creates a bottleneck in the server gateway. Also, there is an interaction between incoming signals and outgoing signals on the gateway during the period when the input signal is overloaded and causes network congestion. However, the **SERV-1** series turns out to be more regular since the data is transferred from the main gateway to thousands of points distributed on the internet, this transfer is simpler compared to the case of incoming traffic.

Since $H(t)$ for the series under study are outside the range $-0.50 \leq H(t) \leq 1.50$, they are very complicated systems that merit independent study to obtain a better description, both quantitative and qualitative.

Notwithstanding the above, the TSA provides valuable information in comparison with the PSA and the DFA allows us to study the behavior of the complex system considered recorded data of traffic flows from and to the internet from an online application server.

## 5 CONCLUSION

In this paper has been presented the application fractal analysis techniques on a time series obtained fram traffic captures coming from an application server connected to the Internet through a high-speed link. The results obtainede show that traffic flow in the dedicated high-speed network link have fractal behavior since the Hurst exponent is in the range of 0.5, 1, the fractal dimension between 1, 1.5, and the correlation coefficient between −0.5, 0. Based on these results, it is ideal to characterize both the singularities of the traffic and its impulsiveness during a fractal analysis of temporal scales.

A detailed analytical study on long-range fractality and dependence for two traffic time series is presented. The time series **SERV-1** and **SERV-2** are examined by three methods: PSA, DFA, and TSA.

It is made clear that there are other techniques to examine LRD that are not addressed in this research, such as dispersion analysis and maximum likelihood estimators.

The main results are summarized as follows:
1) The PSA reports that the series are fractal and have LRD given that the following conditions:
   - $\beta$: $1 < \beta < 2$,
   - $H$: $0.5 < H < 1$,
   - $\rho$: $-0.5 < \rho < 0$, and
   - $D$: $1 < D < 2$.
2) The analysis of fluctuation without trend shows that the series presents the characteristic crossing phenomenon of FP with LRD.
3) The TSA reports that the time series under study, **SERV-1** and **SERV-2**, present a time-dependent Hurst exponent, outside the range (0, 1). Therefore, these time series require an advanced quantitative as well as qualitative description to improve the understanding of the series of internet traffic coming from a high demand environment as it is an online application server, it is recorded that:
   - $H(t)$: $-0.5 \leq H(t) \leq 1.5$,
   - $H$: $0.5 < H < 1.0$, and
   - $D$: $1 < D < 2$.

Finally, fractality and LRD are presented in the studied series that represent traffic captures from a high-speed dedicated link aggregation.